\documentclass[authoryear,preprint,12pt]{elsarticle}

\usepackage{amssymb}

\usepackage{amsmath}
\usepackage{float}
\usepackage{subcaption}

\usepackage{enumitem}

\usepackage{booktabs}
\usepackage{multirow} 

\usepackage{xurl} 
\usepackage[hidelinks]{hyperref}  
\usepackage{csquotes} 

\begin{document}

\begin{frontmatter}


\title{Urban Context and Travel Experience Events: An Exploratory Comparison of Two German Cities}

\author[label1]{Marie Güntert}
\ead[url]{https://orcid.org/0009-0004-7850-3430}

\author[label2]{Esther Bosch\corref{cor1}}
\ead{esther.bosch@dlr.de}
\ead[url]{https://orcid.org/0000-0002-6525-2650}

\cortext[cor1]{Corresponding author}

\affiliation[label1]{organization={Faculty of Engineering and Technology, Furtwangen University},
            addressline=\mbox{Kronenstraße 16},
            city={Tuttlingen},
            postcode={78532}, 
            country={Germany}}
            
\affiliation[label2]{organization={Institute of Transportation Systems, German Aerospace Center (DLR)},
            addressline=\mbox{Lilienthalplatz 7}, 
            city={Braunschweig},
            postcode={38108},
            country={Germany}}

\begin{highlights}
    \item This study compares travel experience of public transportation in an urban and rural area.
    \item There seem to be differences in the impact of influencing events.
    \item Unpunctuality and quality of information are the most influential events in urban Hamburg.
    \item A positive capacity offer and the quality of information influences the most in rural Tuttlingen.
\end{highlights}

\begin{abstract}
The presented study investigates events influencing public transportation experience in both urban (Hamburg) and rural (Tuttlingen) areas in Germany, with the aim of identifying events that affect travel experience and as a result travel behavior. Using a mobile application, 21 participants in Tuttlingen and 70 participants in Hamburg tracked everyday trips, providing real-time evaluations of travel experiences along with situational data. Multi-level regression analyses were applied to assess the impact of events such as punctuality, capacity offer, information about public transportation and others on the ontrip experience. Results indicate that a sufficient public transportation capacity offer has the strongest positive effect in Tuttlingen, whereas a lack of punctuality and low personal well-being have the strongest negative effects. In Hamburg, a lack of punctuality and a negative information event have the largest impacts. These identified effects provide a foundation for decision-making and measures to improve local public transportation.
\end{abstract}


\end{frontmatter}


\section{Introduction}
\label{sec1_Introduction}

The climate crisis is currently one of the biggest social and environmental challenges of the 21st century. While the European Climate Law states the goal of reaching climate neutrality by 2050 and reducing greenhouse gas emissions by at least 55\% in 2030 \citep{eu_regulation_2021}, current projections of established policies and measures indicate a reduction of only 47\% \citep{european_environment_agency_total_2025}. Hence, more ambitious policies and measures in all sectors must be implemented to reach the agreed goal.
The transport sector of the European Union accounted for 21.5\% of emissions in 2023 \citep{environmental_protection_agency_transport_2025}, making it the largest emitting sector in that year \citep{european_environment_agency_greenhousegasses_2025}. Although advances in technology have lead to more emission-efficient vehicles, the number of private vehicles has risen \citep{environmental_protection_agency_transport_2025}. European cities such as Paris, Brussels and Zurich have a pioneering role in effective and sustainable transport policy by passing multiple laws regarding the limitation of vehicles and speed regulation in the city \citep{yannis_review_2024}, motivating residents and tourists into using public transportation as less-emission-emitting alternative to private vehicles \citep{banister_sustainable_2008}. 

Another possible way of influencing peoples' behavior is influencing the travel experience itself. \citet{de_vos_attitude_2021} argue with their travel mode choice cycle that public transportation will be chosen more likely if the travel experience is satisfying. Conversely, unpleasant experiences may lead to dissatisfaction and erode one’s inclination to use transit. More precisely, positive or negative experiences while traveling can shape a passenger’s overall satisfaction with that trip, which in turn influences their attitudes and future choices of transportation mode. For example, \citet{LIM2024201} introduce the refined concept of within-trip subjective experiences, defined as the momentary cognitive and emotional responses to travel situations, and propose that these in-the-moment experiences directly affect travel satisfaction and influence, on the long term, one’s mode-related attitudes and intentions. If a journey is smooth or pleasant, the traveler’s satisfaction will be higher, likely fostering a favorable attitude toward that mode. Therefore, improving the within-trip experience is not just a matter of enhancing a single journey, but may be key to nurturing long-term patronage of sustainable modes. 

\citet{morfoulaki_estimation_2010} define this public transportation satisfaction as the overall experience compared to the associated expectations. It is therefore an important aspect to identify events where an inequality of the experience and expectation changes the satisfaction and choice of travel mode. \citet{van_lierop_what_2018} identified satisfaction influencing events by literature analysis, which revealed that cleanliness, comfort, service, safety and punctuality are the most associated events for causing satisfaction. \citet{susilo_exploring_2014} analyzed a dataset of multiple European city surveys and identified the ease of transfer, the station environment and onboard comfort as general influential events on door-to-door travel satisfaction. They also discovered group-specific differences in event relevance, e.g. safety for female travelers or the absence of hindrances for pedestrians. Furthermore, perceived safety as well as the frequency and reliability of transport services are key predictors for the perceived accessibility of transportation systems\citep{lattman_perceived_2016}, because higher perceived safety reduces stress and uncertainty during travel, while frequent and reliable services enhance flexibility and temporal autonomy. Moreover, the offer or accessibility seems to be closely linked to subjective well-being \citep{de_vos_attitude_2021, olsson_happiness_2013}. Being able to reach everyday activities influences the individual well-being, whereas a lack of mobility accessibility can lead to limited social participation and, consequently, negative evaluation of mobility \citep{de_vos_attitude_2021}. In alignment with the influence of various events, multiple publications have attended the theory of critical incidents as driving influence on travel satisfaction and emotional well-being. These incidents can be defined as particularly satisfying or dissatisfying encounters \citep{allen_effect_2020}. \citet{bitner_critical_1994, friman_structure_2004, gremler_critical_2004} show the immense effect critical incidents have on the remembered travel experience. Taken together, these studies provide important insights into which factors weigh most heavily in shaping overall satisfaction and loyalty to public transport.

While these studies provide valuable insights into satisfaction-influencing events in urban contexts, the role of spatial and structural contexts in shaping these experiences remains under explored. City size and urban density influence the relative attractiveness of transport modes \citep{Liao_2020} and, in turn, travel experience. In dense cities, the time advantage of cars over public transport is often smaller, which encourages greater use of public transport \citep{Liao_2020}. In contrast, in smaller cities and rural areas with longer distances and lower density, cars tend to be much faster and more convenient, reducing the likelihood of choosing public transport. This difference in mobility patterns also contributes to disparities in carbon footprints between urban and rural areas. Higher than average incomes occur mostly in cities and are associated with higher overall emissions \citep{Klein_Taconet_2024}, yet daily mobility in urban areas can be more sustainable due to greater access to efficient public transport systems \citep{uba2020co2alltagsverkehr}. Conversely, in rural areas, the reliance on private vehicles for commuting and daily activities increases greenhouse gas emissions per person. These structural and behavioral contrasts underline the importance of spatial context in assessing both mobility satisfaction and environmental impacts.

With these differences in urban and rural transport the question arises to which extend the affecting events differ depending on the characteristic of the location. Many of the modern solutions for rural mobility require participation of non-dependent people, e.g. carpooling from or to a public transportation mode. As non-dependent people can always use their car, public transportation must archive an attractiveness through satisfaction where it is actively and repeatedly chosen by all residents. There is a focus of literature researching this in the urban context \citep{de_vos_attitude_2021, LIM2024201, susilo_exploring_2014}, however, the rural context remains unexplored. Therefore, the travel experience and the respective affecting events need to be researched as influencing events of choosing public transportation over a private car in rural areas \citep{de_vos_attitude_2021}. Currently, it remains unclear to what extent these urban–rural differences translate into varying intensities of travel experiences and distinct types of events that shape individuals’ evaluations. Insights could prove to be valuable information for planning investments and adapting measures from urban to rural transportation planning.

To explore these questions in different contexts, this study examines two German cities as contrasting cases: Hamburg, a major metropolitan area, and Tuttlingen, a smaller rural town. Comparing these cases can provide initial insights into how urban context may shape travel experiences and their evaluation. Furthermore, Tuttlingen can provide first insights into the influence of events and travel satisfaction in rural areas. Based on these considerations, the following research questions can be defined:
\begin{enumerate}
    \item Which events influence travel experience in Hamburg and Tuttlingen?
    \item Does the influence of these events differ between the cities?
\end{enumerate}

\section{Method}
\label{sec2_Method}

Participants were recruited by using social media channels and via advertising spaces in the public transportation vehicles in Hamburg. This method allowed to recruit regular public transportation users and car commuters to collect a more differentiated opinion. They were then referred to a recruitment questionnaire, which included the contact details and demographic data. Participants had to be 18 years or older and were required to have access to an Android mobile phone.

\subsection{Sample}
\label{subsec1_Sample}
Overall \textit{N} = 21 participants with a mean age of 29.24 years $(SD = 13.55)$ took part in Tuttlingen. Gender distribution was almost balanced, with nine female and eleven male participants One participant chose not to answer this question. 19 participants in Tuttlingen stated using public transportation regularly multiple times a week or daily while three participants use it once a week or once a month. The revised study design, described in \ref{subsec2_StudyProcedure}, resulted in Hamburg in 34 male, 35 female and one non-binary participant $(N = 70)$ with a mean age of $M= 35.93~(SD= 12.96)$. They reported their public transportation usage as follows: 2x once a week, 25x multiple times a week and 43x daily. Written informed consent was obtained from all participants.

\subsection{Study Procedure}
\label{subsec2_StudyProcedure}

This study consisted of two different data acquisition locations: Tuttlingen, a small city in southern Germany, has around 38000 inhabitants and serves as the economic center of its district. Located in the state of Baden-Württemberg on the upper Danube River, Tuttlingen is internationally known as a hub of the medical technology industry, often referred to as the world capital of medical technology. The city is connected to regional and long-distance transport networks, with rail services linking it to larger cities and a local public transport system consisting mainly of buses. The second study location was Hamburg, Germany's second largest city. It has 1.8 million inhabitants and a metropolitan region of around 5.5 million. Located in the country's north on the river Elbe, Hamburg is a major economic and transport hub. The city has a public transport network operated under the Hamburger Verkehrsverbund (HVV), including subway (U-Bahn), suburban rail (S-Bahn), buses, and ferries. 

The study design of this work has been previously tested and further developed \citep{bosch2026shapes, bosch_travel_2025}. With downloading the institute's research app on their Android smartphone, participants were instructed to track six trips with a minimum duration of 15 minutes. During their trip, they were reminded by vibration of their phone to complete a questionnaire about the current situation every five minutes. When they reached their final destination, an additional questionnaire regarding the overall trip was completed. The study was concluded with a retrospective questionnaire of the entire participation. This included stressors, positive and negative aspects of traveling, used apps and the potential willingness to share data for personalized routing. The study manager verified the number of trips and their completeness before paying out money as incentive. Overall participants could get up to 70€ which consisted of 20€ for the first three trips, 30€ after completion of all six trips, 10€ if at least 95\% of questionnaires were answered and ten times 1€ for each additional trip.
This study procedure has been approved by the ethics committee of the German Aerospace Center (reference number 4/25).

\subsection{Collected Data and Measures}
\label{subsec3_CollectedData}
Demographic data was collected though the recruitment questionnaire. The institute's research app recorded public transportation schedule data, actual routes traveled, including the start and end point and the used transport modes. Additionally, questionnaire evaluation data was saved with corresponding pseudonymous login data. The data was prepared before conducting the statistical analysis. First, all data was combined and filtered. For the dataset of Tuttlingen, trips with no start or destination in district of Tuttlingen were sorted out. Evaluations during the trip (called 'ontrip' throughout the paper) were assigned to the associated post-trip evaluation and events. The chosen influencing events in the questionnaire are based on a literature review and identified by analyzing complaints issued to the Hamburg Transport Association as well as qualitative and statistical analysis of previous studies: The ontrip questionnaire contained the following items which were previously identified by using confirmatory factor analysis (manuscript under preparation). Items 1 to 3 were answered with a 5-point Likert-scale:

\begin{enumerate}
    \item The journey at the moment is 1 (miserable) to 5 (excellent).
    \item For achieving my goals, this journey is 1 (a hindrance) to 5 (favorable).
    \item Based on my previous experience with traveling this journey is 1 (below average) to 5 (above average).
\end{enumerate}

Items 4 to 12 could be answered positive or negative, no answer resulted in neutral. Item 13 contained an optional free text answer.

\begin{enumerate}[resume]
    \item Punctuality and transfers, e.g. connection made or transport departed too early (positive | negative, not completed = neutral)
    \item Fellow passengers, e.g. pleasant fellow passengers or disruptive people (positive | negative, not completed = neutral)
    \item Capacity utilization in the transport, e.g. full or plenty of space (positive | negative, not completed = neutral)
    \item Driving behavior, e.g. rough or pleasant driving style (positive | negative, not completed = neutral)
    \item Infrastructure, e.g. clean means of transport or no weather protection available (positive | negative, not completed = neutral)
    \item Information, e.g. alternative routes in case of cancellation or lack of information (positive | negative, not completed = neutral)
    \item Well-being, e.g. nice view, music, temperature, or discomfort, need to hurry (positive | negative, not completed = neutral)
    \item Staff, e.g. nice or unfriendly ticket inspector (positive | negative, not completed = neutral)
    \item Capacity, e.g. good or poor frequency (positive | negative, not completed = neutral)
    \item Other [free text]
    
\end{enumerate}

Collected data was analyzed with multi-level regressions. In Tuttlingen 1876 ontrip evaluations were collected whereas Hamburg resulted in 8188. The mean ontrip experience was defined by calculating the mean value of items 1 to 3 and predicted by the events of items 4 to 12 as independent variables. A random intercept accounted for interpersonal differences of participants. No random slope or interaction effects were modeled. 

\section{Results}
\label{sec3}

\begin{figure}[H]
    \centering
    \includegraphics[width=0.9\linewidth]{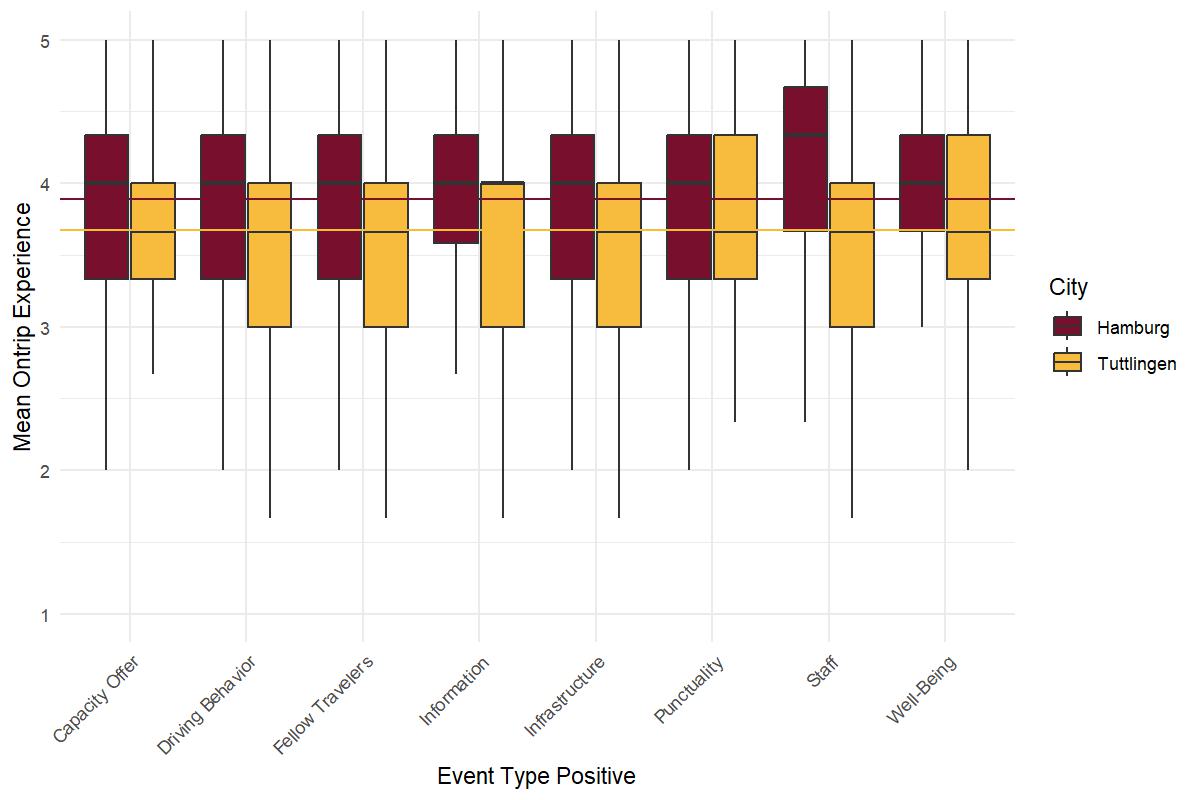}
    \includegraphics[width=0.9\linewidth]{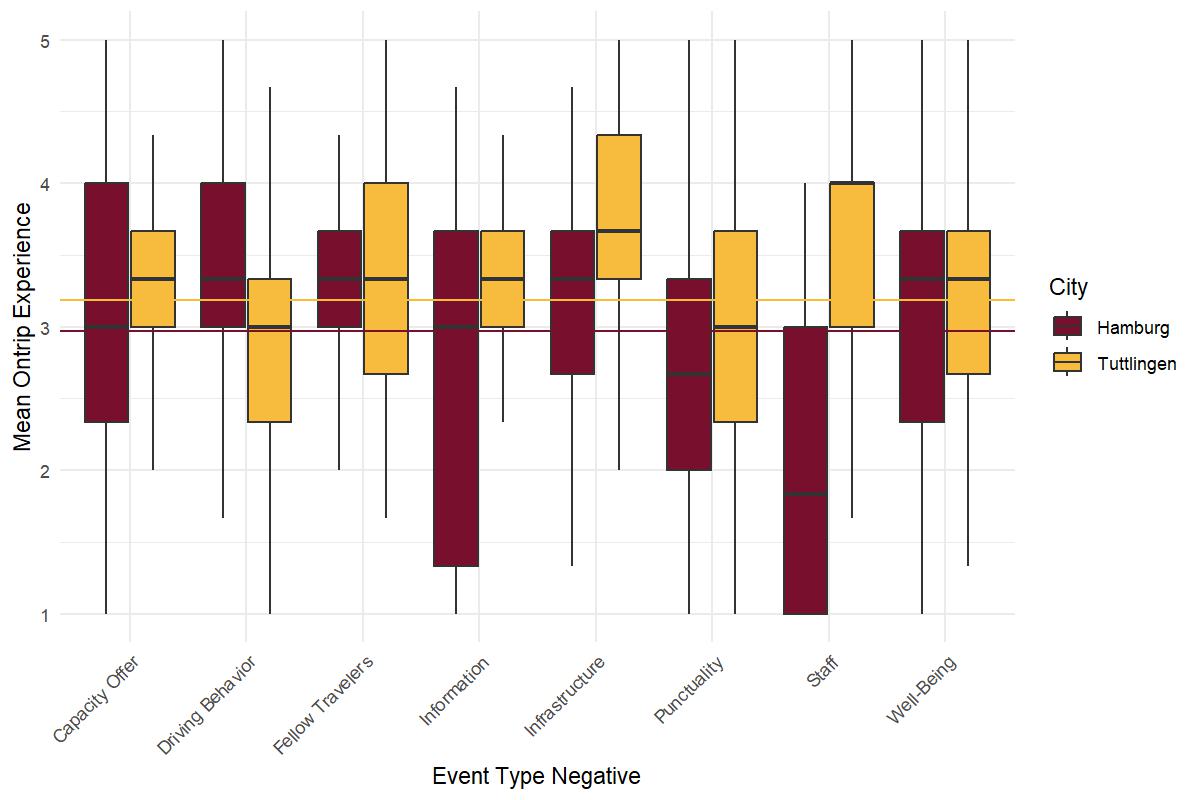}
    \caption{Descriptive visualization of the mean ontrip experience per positive (top) and negative (bottom) event for the cities Hamburg and Tuttlingen. The horizontal lines show the mean value per city for all positive and all negative events. The outliers are hidden.}
    \label{OntripVisualisierung}
\end{figure}

The following chapter presents a visual overview of the data and the results of each regression. The boxplots (Figure \ref{OntripVisualisierung}) show differences in participants’ mean ontrip experience across positive and negative events in Hamburg and Tuttlingen. For the positive events, the interquartile ranges (IQRs) are similar between cities, though Hamburg shows a higher median per event and general mean value, while Tuttlingen’s mean is slightly lower. Negative events display much more variability, with IQRs differing markedly between cities and events, particularly in Hamburg. This is can be seen in comparison of the events of information, capacity, service provision and staff, where medians are notably low, especially with regard to staff and punctuality. Overall, Hamburg has higher mean ratings for positive events, whereas negative events reduce the mean by nearly one point. Tuttlingen, by contrast, shows less difference in the mean ratings between positive and negative events.

Multi-level regression models with an AR(1) correlation structure were fitted to predict mean ontrip experience for each city. Residuals of the initial model without autocorrelation show strong positive autocorrelation ($Durbin- Watson = 0.26, p \leq .001$), so a first-order autoregressive structure ($AR(1) \phi = 0.90$) was included. ACF and PACF plots of normalized residuals indicated that AR(1) adequately captures temporal dependence. A series of model comparisons was conducted using likelihood-ratio tests and information criteria (AIC and BIC). If the likelihood-ratio test was non-significant, the more conservative model was being preferred, particularly when supported by BIC. This approach was chosen due to the large sample size and the risk of detecting statistically significant but practically negligible effects. Table \ref{results_tuttlingen} presents the results of the fitted model, which showed the lowest AIC (821.08) and BIC (920.75) for the compared models with the excluded event \textit{staff}. Including all collected events resulted in a slightly worse model fit ($\Delta \mathrm{BIC} = +14.20$) with the likelihood-ratio being non-significant, $x^2(1)= 0.87, p=.646$.

Non-significant events include negative driving behavior, positive fellow travelers, positive personal well-being and negative infrastructure. All negative coefficients are associated with lower experience ratings, while positive coefficients are associated with higher ratings. The strongest effects in Tuttlingen are caused by a positive capacity offer ($\beta=0.31$), negative punctuality ($\beta=-0.26$), negative well-being ($\beta= -0.23$) as well as negative ($\beta=-0.22$) and positive ($\beta=-0.23$) information.

\begin{table}[H]
\centering
\caption{Results of multi-level regression for Tuttlingen, significant events are highlighted. Neutral served as reference for the categorical event variables. $df=1842$}
\label{results_tuttlingen}
\begin{tabular}{@{}llrrr@{}}
\toprule
\textbf{Event} & \textbf{} & \multicolumn{1}{l}{\textit{\textbf{$\beta$}}} & \multicolumn{1}{l}{\textit{\textbf{t}}} & \multicolumn{1}{l}{\textit{\textbf{p}}} \\ \midrule
Intercept   &   & 3.17  & 28.33  & $\leq$ \textbf{.001}  \\
\multirow{2}{*}{Driving behavior} & negative  & 0.07  & 1.13 & .258                           \\
                                            & positive  & 0.18  & 4.46 & $\leq$ \textbf{.001}                          \\
\multirow{2}{*}{Information}                & negative  & - 0.22 & - 3.67  & $\leq$ \textbf{.001}                         \\
                                            & positive  & - 0.23  & - 4.44 & $\leq$ \textbf{.001} \\
\multirow{2}{*}{Fellow travelers}          & negative  & - 0.17  & - 3.78 & $\leq$ \textbf{.001}                          \\
                                            & positive  & 0.06   & 1.54  & .123                           \\
\multirow{2}{*}{Personal well-being}         & negative  & - 0.23 & - 5.14 & $\leq$ \textbf{.001}                          \\
                                            & positive  & 0.06  & 1.66  & .097                           \\
\multirow{2}{*}{Capacity offer}             & negative  & - 0.09  & - 2.02  & \textbf{.044}                                   \\
                                            & positive  & 0.31 & 6.15 & $\leq$ \textbf{.001}                          \\
\multirow{2}{*}{Punctuality}                & negative  & - 0.26 & - 5.26 & $\leq$ \textbf{.001}                          \\
                                            & positive  & 0.17  & 3.67  & $\leq$ \textbf{.001}                          \\
\multirow{2}{*}{Infrastructure}     & negative  & 0.09  & 1.74  & .082                           \\
                                            & positive  & 0.20  & 4.21  & $\leq$ \textbf{.001}                         \\ \bottomrule
\end{tabular}
\end{table}

The model for Hamburg includes all event-types with an $\mathrm{AIC} = 2088.50$ and $\mathrm{BIC} = 2228.71$ and uses an AR(1) correction as well. Excluding the event \textit{staff}  would result in a worse model fit ($\Delta \mathrm{BIC}=4.95$) with a significant likelihood-ratio, $\chi^2(1)= 22.44, p\leq .001$.

\begin{table}[H]
\centering
\caption{Results of multi-level regression for Hamburg, significant events are highlighted. Neutral served as reference for the categorical event variables. $df= 8101$}
\label{results_hamburg}
\begin{tabular}{@{}llrrr@{}}
\toprule
\textbf{Event} & \textbf{} & \multicolumn{1}{l}{\textit{\textbf{$\beta$}}} &  \multicolumn{1}{l}{\textit{\textbf{t}}} & \multicolumn{1}{l}{\textit{\textbf{p}}} \\ \midrule
Intercept                                   & & 3.36  & 76.01 & $\leq$ \textbf{.001}                          \\
\multirow{2}{*}{Driving behavior} & negative & - 0.07  & - 2.33  & \textbf{.020}                                    \\
                                            & positive & 0.13 & 7.75 & $\leq$ \textbf{.001}                          \\
\multirow{2}{*}{Information}                & negative & - 0.30 & - 11.34 & $\leq$ \textbf{.001}                          \\
                                            & positive & 0.08  & 4.26 & $\leq$ \textbf{.001}                         \\
\multirow{2}{*}{Fellow travelers}          & negative  & - 0.13 & - 7.81 & $\leq$ \textbf{.001}                          \\
                                            & positive & 0.07  & 4.20 & $\leq$ \textbf{.001}                         \\
\multirow{2}{*}{Personal well-being}         & negative & - 0.13 & - 7.21  & $\leq$ \textbf{.001}                          \\
                                            & positive & 0.15  & 8.34 & $\leq$ \textbf{.001}                          \\
\multirow{2}{*}{Capacity offer}             & negative & - 0.15 & - 6.34 & $\leq$ \textbf{.001}                          \\
                                            & positive & 0.14 & 7.36 & $\leq$ \textbf{.001}                          \\
\multirow{2}{*}{Staff}                & negative  & - 0.15 & - 2.66 & 	\textbf{.008}                          \\
                                            & positive  & 0.07  & 3.75  & $\leq$ \textbf{.001}              \\
\multirow{2}{*}{Punctuality}                & negative  & - 0.50  & - 21.38  & $\leq$ \textbf{.001}                          \\
                                            & positive & 0.07 & 3.73  & $\leq$ \textbf{.001}                         \\
\multirow{2}{*}{Infrastructure}     & negative & - 0.07 & - 2.79  & \textbf{.005}                                    \\
                                            & positive & 0.12 & 6.21  & $\leq$ \textbf{.001}                          \\ \bottomrule
\end{tabular}
\end{table}

\section{Discussion}
\label{sec4_disc}

This study focuses on analyzing how within-trip events shape the public transportation experience in two contrasting spatial contexts. To address research question 1, a statistical analysis was conducted to examine the influence of the defined events on the mean on-trip experience.

For Tuttlingen the different coefficients can be interpreted as follows: A positive capacity offer seems to have the highest positive influence $(\beta=0.31)$ on the mean ontrip experience and could be explained by the rural nature of the district Tuttlingen, where sufficient or above-average experienced space is perhaps perceived and evaluated more positively than in urban areas.
Interestingly, both the absence of information, as well as the presence of information, have a negative impact. This phenomenon can be attributed to the quality of the displayed information, which suggests that the information received by the participants was generally perceived as unsatisfactory or inadequate.
It is noteworthy that positive driving behavior and infrastructure significantly enhance the experience, while negative driving behavior or negative infrastructure have a negligible impact. Again, this could be attributed to the rural characteristic of Tuttlingen, where participants may not anticipate adequate behavior or infrastructure and consequently experience positive surprise when such qualities are present. Alternatively, a high widespread appreciation regarding public transportation could explain this effect as well.
In contrast to this, fellow travelers and personal well-being have a sorely negative influence. This could indicate that in a public transport system with low overall usage, individual negative experiences become particularly salient, while positive or neutral behavior is hardly noticed.

The most influential event in Hamburg is a negative punctuality $(\beta= -0.50)$. As metropolitan areas offer a tight transportation schedule, one could assume that individuals may prioritize the scheduling of their travels with minimal allowance for contingencies, consequently leading to delays in their travel plans once a transportation mode becomes unpunctual. This is surprising as the public transportation in metropolitan areas offers a high flexibility due to their tight schedule and many transport modes. However, the same argument could explain the effect size of a negative information event $(\beta= -0.30)$. In such situations, a fast alternative needs to be offered to keep the delay low. Without information or with poorly rated information, this rescheduling could be inefficient and cause stress.
The strongest positively influencing event is the personal well-being $(\beta=0.15)$, which is relatively low compared to the negative events of punctuality and information. It seems that the travel experience can be positively influenced the highest, not by improving the public transportation, but by improving the individuals' well-being. This would support the previously reported influence of the personal well-being \citep{de_vos_attitude_2021}. However, this should be investigated further, as the difference to the other events' effect sizes (ranging from $\beta=0.07$ to $\beta=0.14$) is low and the data of Tuttlingen does not support this effect. 

Research question 2 addresses the differences between the cities. With an intercept value of 3.17 for the mean ontrip experience in Tuttlingen, Hamburg has a slightly higher value of 3.36. This could potentially indicate a better overall experience in the public transportation of Hamburg, however, both values are relatively high and display only a marginal difference.
The event \textit{staff} provided in Tuttlingen, in contrast to Hamburg, no significant and BIC-related improvement of the statistical model for Tuttlingen and should be therefore verified of its importance. It is possible that the low public transport offer of Tuttlingen leads to fewer staff presence. Due to the rural nature, it should be investigated further which additional events have an influential impact. This could also be the reason why negative punctuality has the highest effect in Hamburg, whereas a positive capacity offer influences the strongest in Tuttlingen. Interestingly, the most influential events are the same in both cities. Although the order and intensity varies, punctuality, capacity, well-being and information are the important events, which could indicate city-unspecific, general influences.
These differences between the cities highlight the importance of contextual events, including urban density and transport system characteristics, whereas the similarities show the fundamental problems of public transportation in Germany. 

The previous discussed findings for Tuttlingen could be interpreted as a higher acceptance of errors in this rural area, since negative driving behavior and a negative infrastructure have a marginal effect size $(\beta_{driving}=0.07, \beta_{infrastructure}=0.09)$. Additionally, the ontrip experience shows a smaller difference between the all positive and all negative events mean value $(\Delta=0.49)$ than in Hamburg $(\Delta=0.92)$. This interpretation would strengthen the argument that currently mostly dependent people use rural public transportation, as they have no alternative and therefore no choice but to accept certain conditions to avoid transport related social exclusion \citep{Lieszkovszky_2018}.

It has been demonstrated that all identified events based on previous work \citep{bosch_travel_2025, bosch2026shapes} and literature \citep{morfoulaki_estimation_2010, van_lierop_what_2018, allen_effect_2020} do indeed have an influence in Hamburg, thus strengthening the findings about positive and negative critical incidents \citep{bitner_critical_1994, friman_structure_2004, gremler_critical_2004}. However, with the dataset including 8188 data points, the significance could be a consequence of the large sample size and some effect sizes are relatively small. Most of the defined events influence the experience in the rural Tuttlingen. However, further identified aspects such as frequency and reliability \citep{lattman_perceived_2016}, the ease of transfer or group-specific events \citep{susilo_exploring_2014} should be reviewed regarding their influence in rural areas.

Previously identified highly positive or negative experienced events or critical incidents influence the travel satisfaction and can as a result change the probability of showing the sustainable behavior \citep{de_vos_attitude_2021} of using public transportation instead of private vehicles. By evaluating the experience of people towards public transportation and identifying influencing events shaping the experience of current trips or satisfaction of previous trips, improvements of these events could result in a more positive attitude and therefore a higher probability of using public transportation as sustainable behavior \citep{bosch_travel_2025}. This behavior would reduce the use of individual transportation and, consequently, lower the amount of emissions produced by the European transport sector.

Conclusively, the influencing events for major cities are identified and could now be validated with this study. Following the travel mode choice cycle \citep{de_vos_attitude_2021}, a next step would be to define improvements of measures for these events and evaluate the resulting probability of choosing public transportation as sustainable behavior with the respective satisfaction.
These results provide a basis for influencing sustainable and inclusive travel behavior and therefore the carbon footprint of individuals. They can be seen as a small impulse for decision-making and drafting policies supporting the implementation of the European Climate Law.

\section{Limitations}
\label{sec5_lim}

With every study limitations arise. Firstly, the rural characteristic of Tuttlingen made it difficult to recruit as many participants as in Hamburg. This is simply because not every resident has access to public transportation and is reliant on a personal car or car sharing. Together with the smaller population it results in a small sample size which further influences the statistical power and comparability of the dataset to Hamburg. Additionally, 19 of the 21 participants used public transportation in Tuttlingen regularly. This poses the question to which extend the targeted user group, car commuters, is represented by the sample and whether these results are applicable for influencing travel behavior towards public transportation of people who currently use public transportation only to a limited extent or not at all.
Secondly, with a mean age of 29.24 and 35.93 the samples are rather young. Whether this is due to a younger population being more dependent on public transportation or because the usage of a phone application resulted in a technical barrier for elderly residents is currently unclear. However, an older population might be influenced in a different way by the identified events.

Another potential limitation of the present analysis is the potentially inconsistent interpretation of individual event categories by participants. This is especially evident in the context of the information event, where the presence or absence of information, or its perceived quality, can be viewed as positive and negative which in turn influences the data in a contrary way. In future surveys, this should be reduced through the implementation of more precise items.

While prior evidence suggests that momentary measurement generally does not distort overall evaluations \citep{heron2013_intensive}, some studies indicate that frequent reporting may amplify the salience of momentary fluctuations \citep{ariely_combining_1998}. That effect does not necessarily has to be suppressed with an adapted study design, as it could rise the awareness of travelers using public transportation and result in a more reflected experience. However, it needs to be investigated in order to be able to answer the influence on the results with certainty. Furthermore, the retrospective evaluation might influence the satisfaction, as suggested by the peak-end rule of \citet{kahneman_when_1993}. Additional analysis of the collected post-trip data could clarify, whether the experience could be improved with measures after the trip and to what extend.

Lastly, the explanations for the results discussed in chapter \ref{sec4_disc} are based on the statistical analysis of the data. No qualitative information served as foundation which should be subject to future analysis.

\section{Conclusion}

This study set out to investigate how within-trip events shape the public transportation experience in two contrasting spatial contexts, an urban and a rural setting. Building on the premise that travel satisfaction plays a crucial role in shaping long-term travel mode choice, the analysis provides empirical evidence that momentary events are key drivers of overall travel evaluation in both contexts.

Across both Hamburg and Tuttlingen, core factors such as punctuality, information, capacity, and personal well-being emerged as central determinants of on-trip experience. This indicates that, at a fundamental level, similar mechanisms underlie travel satisfaction regardless of spatial context. However, the interpretation of these effects differs notably between the two cases. While the lack of punctuality (negative punctuality) and insufficient information dominate the urban experience in Hamburg, the rural context of Tuttlingen highlights the importance of capacity availability and suggests a comparatively higher tolerance toward negative incidents. These findings point to the relevance of contextual conditions.

The results therefore underline that transport policies and service improvements cannot be transferred one-to-one from urban to rural settings. Instead, they need to account for structural differences in infrastructure, service provision, and user expectations. In particular, the findings suggest that rural public transportation must not only ensure basic accessibility, but also actively enhance the travel experience to remain an attractive option, especially in light of the higher carbon footprint of the rural population.

At the same time, this study highlights a remaining research gap. While urban travel behavior and satisfaction have been widely studied, empirical evidence for rural contexts remains limited. The observed differences in event relevance and effect sizes indicate that further research is needed to disentangle which influences are inherently linked to spatial characteristics and which are driven by other factors such as user groups or service quality.

In conclusion, improving public transportation experience represents a promising lever for promoting sustainable mobility behavior. By identifying and addressing the most influential events in different spatial contexts, transport planning can contribute not only to increased user satisfaction but also to broader objectives such as social inclusion and climate change mitigation.


\section{Declaring of generative AI and AI-assisted technologies in the writing process}
\label{sec6_AI}

Statement: During the preparation of this work, the authors used ChatGPT to improve wording and optimize the readability of the text. DeepL was used for the translation of individual terms and passages and checked against specialist literature. DeepL Write was used to refine stylistic and linguistic nuances. Zotero was used to manage and organize the literature sources used in all chapters including the references. After using these tools, the authors reviewed and edited the content as needed and takes full responsibility for the content of the published article.

\section{Declaration of Interest.}
The authors declare no conflict of interest.

\section{Acknowledgements.}
The authors thank Anke Sauerländer-Biebl for enabling the data collection by providing technical support with data collection. We also thank Jakob Dietze for his support with the data acquisition.

CRediT roles: 
Conceptualization, Data curation, Formal analysis: MG and EB; Funding acquisition: EB; 
Investigation, Methodology: MG and EB; Project administration: EB; Validation: MG and EB; 
Visualization: MG; Writing – original draft: MG; Writing – review and editing: MG and EB; Supervision: EB.

Funding: This work was supported by mFUND of the German Federal Ministry of Transportation under the project 'Erlebensatlas' \footnote{https://www.dlr.de/en/ts/research-transfer/projects/erlebensatlas}, grant number 19F1197A, and by the Hamburger Hochbahn AG.


\bibliographystyle{elsarticle-harv} 
\bibliography{02_references}

\end{document}